# How Large is too Large? A Review of the Issues related to Sample Size Requirements of Regional Household Travel Surveys with a Case Study on the Greater Toronto and Hamilton Area (GTHA)

Khandker Nurul Habib, Ph.D., P.Eng.
Percy Edward Hart Professor in Civil & Mineral Engineering
University of Toronto
khandker.nurulhabib@utoronto.ca

Wafic El-Assi, M.A.Sc.
Department of Civil & Mineral Engineering
University of Toronto
wafic.el.assi@mail.utoronto.ca

Tian Lin, M.Eng
Department of Civil & Mineral Engineering
University of Toronto
tian.lin@mail.utoronto.ca

**Abstract**

The paper presents a review of sample size issues related to regional household travel surveys. A review of current practices reveals that different perspectives and, as a result, different practices exist in Canada, US, and abroad on sample size. The paper uses data from the Transportation Tomorrow Survey (TTS) - a household travel survey conducted every five years in the Greater Toronto and Hamilton Area (GTHA) - for a set of empirical investigations that asses the adequacy of household travel survey samples. The empirical investigations reveal that even with a 5% sample size, a full representation of the population and its corresponding travel behaviour may be difficult (at the 95% confidence level). Therefore, based on the results of the empirical investigations and the literature review, the paper proposes a flexible framework for household travel survey sample size determination, especially for Canadian municipalities. The findings and recommendations of the paper are unique in transportation literature as they shed light on the statistical adequacy of household travel surveys' sample size – an issue that lacks consensus in today's practice of survey design. Further, the paper contributes to the literature by presenting a systematic review and recommendations on sample size determinants of regional household travel surveys.

# 1. Introduction

Since the 1980s, trip diaries of household members have been collected as part of household travel surveys (Harvey 2003). Practitioners have always had issues with these surveys, especially with data quality and low response rates. There have been numerous efforts to improve household surveys, most of which are concerned with reducing missing/omitted trip information and response burdens to reduce non-response rates. However, the reduction of non-response rate may not be correlated with such endeavours. Non-response rate is germane to evolving lifestyle, technology reliance and increasing time pressure of modern urban life. Considering the aforementioned factors, Stopher and Greaves (2007a) suspected that future household travel surveys would not be restricted to their existing form of diaries, but offered no alternative. They predicted that household travel surveys, especially in the form of travel diaries, would continue to be the only reliable passenger travel data for urban transportation planning in the foreseeable future. The use of GPS, Smart phones, panel surveys, continuous surveys and other innovative approaches are recommended with caution. However, with the exception of continuous surveys, it is often expected that such advanced approaches would complement the core cross-sectional household travel survey rather than replace it.

The household travel survey provides basic information on household and individual level characteristics, and activity-travel information of household members that are fundamental to the development of any comprehensive travel demand model (Goulias 2013). However, sample size determination for household travel surveys has proven to be a controversial element in the urban planning process, as statistical requirements are often dominated by cost and political considerations (NCHRP 2008). As a result, there is no consensus on sample sizes for household travel surveys in practice. Further, despite plenty of valid reasons for switching to continuous surveys (Ampt and Ortuzar 2004; Ortuzar et al 2011), the use of large cross-sectional surveys remains dominant across the major metropolitan areas in North America and abroad. Nonetheless, examples of empirical investigation on the adequacy of different sample sizes of cross-sectional household travel surveys are hard to come by.

Almost all travel survey researchers recommend a combination of data sources to replace large cross-sectional travel surveys. These data sources include small sample panel surveys with application of GPS/Smart phone, continuous cross-sectional surveys as opposed to simple cross-sectional surveys, etc. Data fusion is considered to be the statistical tool to combine all such datasets to produce a core database, synonymous to the large scale household travel survey. One of the key arguments for replacing large sample cross-sectional surveys by continuous, panel or repeated cross-sectional surveys is the lower sample size requirement. If the rolling average of aggregate travel information is considered (e.g. trip rates, modal share, etc.), a smaller repeated cross-sectional/continuous travel survey can provide data of similar statistical strengths to that of a once-in-a-while large cross-sectional survey. Some regions (e.g. Calgary, Montreal) in Canada have been testing the feasibility of replacing large sample household travel surveys by continuous repeated cross-sectional surveys. Although countries like Australia (e.g. the Sydney Household Travel Survey) have adopted the continuous survey method as their main approach to conduct household travel surveys, there is no firm evidence that this is a practical alternative to serve the passenger travel data needs of large urban areas. Further, continuous surveys have been reported to have a number of drawbacks such as poor data quality and high staff turnover (Ampt

& Ortuzar, 2011). In either case, the sample size of a household travel survey remains a critical element for urban travel demand researchers.

This paper investigates the issue of sample size requirements for household travel surveys from the perspective of adequate data availability to maintain data-driven and evidence-based planning in large metropolitan areas. The objective of the paper is to highlight the sample size requirements for large household travel surveys through focusing on the sample size requirements of trip generation and, more importantly, trip distribution. The paper is inspired by the prospect of re-designing one of the oldest and most regular (every 5 years since 1986) household travel surveys in North America, the Transportation Tomorrow Survey (TTS) of the GTHA (DMG 2015).

The sample size of the TTS has been traditionally around 5% of the GTHA household population. The TTS started in 1986, and the latest cycle ($5^{th}$) was in 2011-2012. The TTS sampling frame leverages a land-line telephone directory to conduct telephone interviews with prospective respondents. In its latest cycle (2011-2012), a web version of the telephone interview was introduced as an alternative option to respondents. The TTS, however, is now facing issues concerning the under coverage of certain population cohorts from the use of the land-line telephone directory as a sample fame. This has resulted in the under representation of key population segments, despite the survey's large sample size. This also prompts the issue of sample size adequacy. Therefore, this paper focuses on sample size requirements of large household travel surveys.

Contrasts in practice of household travel surveys in Canada and abroad were drawn from existing literature. An empirical investigation on the representativeness of past TTS surveys was carried out. The empirical investigation was further extended to examine the sample size necessary to capture trip distribution. Finally, a sample size scheme proposal for the TTS, which may also be relevant for similar regions around the world, is presented.

The paper is organized as follows. Section 2 presents a literature review on household travel survey's sample size requirements. Section 3 discusses the differences in practice of household travel survey sample size determination in Canada and abroad. Section 4 presents an empirical investigation on the representativeness of a large scale household travel survey in Canada - the TTS. Section 5 presents an examination of sample size requirements for trip distribution using graphical methods. Section 6 presents a recommendation for the determination of adequate sample sizes for household travel surveys. Section 7 and 8 present an old-new methodology to calculate sample size, and the associated sampling rate for augment samples for the core of household travel surveys. The paper concludes with a summary of key findings, limitations and recommendations for further research.

## 2. Review of Existing Literature on Household Travel Survey Sample Size

Statistical procedures for estimating sample size of different variables of interest are well-established. Kish (1965), Richardson et al (1995) and the NCHRP report (2008) are a few examples of many other similar sources that explain systematic approaches for estimating sample sizes considering specific objective variables of measurement. They illustrate sample size determination processes for random and stratified random sampling, along with various other combinations of methods for both continuous and discrete variables. Hence, reviewing the processes for sample size determination is not under debate by any means. The question that is yet to be answered is: what is the most appropriate sample size and, consequently, sampling rate for a multi-objective household travel survey for large urban areas? We are referring to household travel surveys that are conducted by one or multiple planning agencies of an urban area to collect data necessary to drive evidence-based planning processes. It is worth noting that the use of sampling rate will be repeated in this paper to illustrate the relationship between sample size and the overall population of the survey area. Further, sampling rate is helpful in reflecting sample size requirements for trip distribution as a function of total trips conducted between O-D pairs. Indeed, sampling rate is merely the sample size divided by a population.

The dilemma of household travel survey sample size determination has historical breaks. Sample size requirements of household travel surveys have been a concern for transportation planners since the 1970s and early 1980s (Stopher and Meyburg 1979). It has been established that the sample size of a household travel survey is dependent on the purpose of the survey, population representation, variability of key variables that are measured in the survey (e.g. trip generation rate, trip length distribution, trip distribution patterns, modal shares etc.), allowable tolerance of errors in measurement, and the desired confidence limit on the estimates from the sample. After more than a decade-long pause, the next phase of research on this topic showed up in the mid-1990s. Interestingly, during this time period (in the 1980s and early 1990s), the concept of travel demand has undergone a paradigm shift from aggregate trip-based approach to the disaggregate tour or activity-based approach.

Earlier studies note that household travel surveys used to be 5% to 10% of population size (Smith 1979). However, Smith (1979) argued that if the main purpose of household travel surveys is to develop travel demand models, the re-generation of such large surveys at regular time intervals is redundant. If stable estimates of key variables from previously conducted large surveys are available, a small sample size may be sufficient for updating the different components of a travel demand modelling system. An empirical investigation proposed by Smith reveals that, with proper estimates of mean and variances of key variables from a large scale survey, a sample of less than 1,200 households may be enough for updating a cross-classification model of trip generation as a function of automobile ownership and income. However, if trip rates per jurisdiction (i.e. zone/county) of a multi-jurisdiction study area are of concern, a sample of 1,100 households per jurisdiction is necessary. Smith proposed a systematic procedure for estimating the sample size of a small scale household travel survey necessary for developing the various travel demand modeling components. Smith identified trip distribution as the critical element that drives up the sample size requirement of household travel surveys. He proved that a 4% sample is necessary to achieve a 90% confidence interval with a 25% standard error for trip interchanges between Origin-Destination pairs with less than 1,100 trips in between. The confidence interval of a sample provides the likelihood that the selected interval encompasses the true value of the population variable (e.g. number of trips between an origin-destination pair) (Fleiss 2003). This has led academicians such as Ortuzar to suggest the use of secondary data sources (e.g. cordon counts, etc.) to create and update Origin-Destination matrices as opposed to conducting a household travel survey of a relatively large sample size.

However, Smith considered pure random sampling for sample size calculation, which is not usually considered to be the sampling method of choice for large study areas with multiple jurisdictions/municipalities. Instead, geographic stratification tends to be the method of choice for most large-scale household travel surveys. Demographics, travel patterns and behavior are also considered in sample stratification and survey design of large scale household travel surveys. Stopher (1982) extended the proposed procedure of Smith for stratified random sampling. However, his sample size calculation considers that accurate estimates of mean and variance of key variables would be available. Nonetheless, the availability of such input statistics for stratified geographic areas is difficult to assume. For example, Kollo and Purvis (1984) collected household travel survey data over a 20-year period and found that trip rates only remain stable over time if aggregated. In other words, disaggregation of trip rates by purpose causes instability over time.

The next remarkable document that has, in part, focused on household travel survey sample size determination is TMIP (1996). The report states that sample rates for household travel surveys are the result of a trade-off between budgetary constraints and sample size requirements for accurate representation of the sampled population. It also reports that the exhaustive objectives of household travel surveys inhibit the optimization of sample size estimation (i.e. too many important variables). Further, the document recommends that one out of every hundred households (1% of the population) for large urban areas, and one of every ten households (10%) for small suburban areas should be the minimum sample size for household travel surveys. The report capitalizes on the fact that the drop of household travel surveys' sample sizes from over 4% to less than 1% of households happened during the late 1980s without necessarily affecting the accuracy of demand modelling. This is another evidence of the impact of the research conducted by Smith (1979) and subsequent researchers. On the other hand, it also recognizes the importance of large sample sizes for increasing the reliability of sample statistics. It provides a step-by-step procedure for sample size estimation of various types of target variables, and for different sampling procedures. However, it provides no definite guideline for sample size determination for a generalized multi-objective household travel survey that can be used by different planning agencies for various purposes.

Greaves and Stopher (2000) highlighted the importance of large sample household travel surveys while recognizing the increasing cost of collecting larger sample sizes. Large sample sizes are being increasingly demanded for developing advanced disaggregate travel demand models. The authors proposed a simulation technique to generate synthetic household travel surveys in the absence of large sample household travel surveys. Simulation takes the conditional distributions from the National Personal Travel Survey (NPTS) and Public Use Microdata Sample (PUMS) to generate an artificial sample. The PUMS is of a 5% sampling rate and so is considered a reliable data source. Pointer et al (2004) also used the same procedure to generate synthetic household travel survey data for Sydney. They used the Sydney household travel survey, a relatively small continuous survey of 3,000 households per year. They pointed out that, though estimating a travel demand model for a region may not need a large household travel survey, portraying an accurate picture of the spatial distribution of travel demand within the region requires a large sample size.

Ampt and Ortuzar (2004) presented a comprehensive discussion on the sample size requirements of household travel surveys. The authors investigated the sample size of Origin-Destination (O-D) trips from a group of only 34 zones in Santiago by using data from the 1991 Santiago O-D survey. They re-confirm that they would need at least a 4% sample to achieve a 90% confidence and 25% standard error for the number of trips between O-D pairs if they were to conform to Smith's (1979) proposition. A 4% sample size was identified as too large considering trip distribution as a meagre objective of the overall household travel survey. They proposed an alternative heuristic algorithm based on stratified random sampling of selected socio-economic variables. However, they also recognized the fact that actual sample size requirements may be very large if geographic distributions of key variables (e.g. zonal or sub-regional estimates of household car ownership) are of concern. The authors also propose that large metropolitan areas should implement small sample continuous household travel surveys with once in a while large sample cross-sectional surveys. Stopher and Greaves (2007b) further proved that if a panel survey is to be the method of choice, sample size requirements reduce drastically. The combination of one of the aforementioned approaches with the use of GPS devices, and weeklong surveys instead of a one-day survey is capable of further reducing sample size requirements for household travel surveys (2007a).

In addition, Stopher et al (2008) proved that even with increasing response burden and the possibility of attrition, a week-long household travel survey can be more efficient than a 24-hour travel survey. A week-long survey demands a smaller sample size requirement. It also provides a rich dataset that can reflect the dynamics of travel behaviour. As an empirical anecdote, the authors proved that a 7-day GPS assisted household travel survey would require a sample size that is 35% less than that of a typical 1-day household travel survey. Similarly, Bolbol et al (2012) suggested a procedure for estimating sample size requirement for GPS assisted household travel surveys. They suggest that the temporal variability of travel mode choices has to be carefully considered for sample size determination. Further, Goulias et al (2013) considered a week-long GPS assisted household travel survey as the core for their core-satellite approach of urban travel data collection. They recommend small yet detailed household travel surveys as the core, which should follow the form of week-long travel diaries of household members. However, the small sample has then to be complemented by a series of carefully designed satellite (synonymous to an augment survey) surveys targeting specific variables that are under or unrepresented in the core. Nevertheless, their proposal provides no guidelines on sample size requirements.

The NCHRP report (2008) stated that even strictly designed (statistically efficient) sample sizes may not be sufficient for serving many of the critical objectives. The 1990 Southern California household travel survey was presented as a case study. A statistically adequate sample size was estimated (3,500 to 5,000 households). However, the actual sample size was selected to be 15,000 households, partly due to political reasons. Interestingly, even with such a large sample size, data was not adequate. Low transit modal shares proved to be a major problem, resulting in a small number of observed transit trips. The number of trips was not large enough to estimate a reasonable mode choice model. Therefore, NCHRP (2008) suggested that the sample sizes of household travel surveys should be based on proper stratification of the key variables of concern (socio-economic variables, modal shares, etc.). It also suggested that, as an alternative to larger sample sizes, a designed sample size should be complemented by augment samples collected for certain zones/sub-regions with a small number of observations.

In summary, it is evident that there is a lack of consensus on the appropriate guidelines for establishing sampling rates for household travel surveys. Although theoretically the sample size can be quite low, the actual sample sizes of urban household travel surveys vary widely. Different trends are observed in different parts of the world. The following section presents a discussion on this.

## 3. Comparison of Recent Household Sampling Rates from Around the World

Table 1 presents a list of recent household travel surveys from the US, Canada, Australia, Europe and South America. The selection of this list is based on web-accessible information. Although it does not provide an exhaustive list of all household travel surveys around the world, it portrays the distinctive approaches in major cities/urban regions.

**Table 1: Sample Sizes of Recent Household Travel Surveys Around the World**

| City/Region | Survey | Year | Sampling Rate (households) |
|---|---|---|---|
| **Canada** | | | |
| Calgary | Calgary Travel and Activity Survey[1] | 2012 | 3.4% |
| Edmonton | Edmonton Household Travel Survey[2] | 2005 | 2.6% |
| Greater Montreal Region | Greater Montreal Area Origin-Destination Survey[3] | 2013 | 4.6% |
| Greater Toronto and Hamilton Area: GTHA | Transportation Tomorrow Survey: TTS[4] | 2011-2012 | 5.0% |
| National Capital Region: NCR | NCR Origin-Destination Survey[5] | 2011 | 5.0% |
| Saskatchewan | Saskatoon Household Travel Survey[6] | 2013 | 3.0% |
| Vancouver | Metro Vancouver Regional Trip Diary Survey[7] | 2011 | 2.2% |
| Winnipeg | Winnipeg Area Travel Survey[8] | 2007 | 3.3% |
| **United States** | | | |
| Atlanta Region | Regional Travel Survey[9] | 2011 | 0.5% |
| Chicago Metropolitan Area | Regional Household Travel Inventory[10] | 2007-2008 | 0.44% |
| Dallas Metropolitan Area | Household Travel Survey[12] | 2008 | 0.24% |

---

[1] http://wwwsptest.calgary.ca/Transportation/TP/Pages/Planning/Forecasting/Forecasting-surveys.aspx
[2] http://www.edmonton.ca/transportation/RoadsTraffic/2005_HTS_Region_Report_FINAL_Oct24_06.pdf
[3] https://www.amt.qc.ca/fr/a-propos/portrait-mobilite/enquetes-en-cours
[4] http://www.dmg.utoronto.ca/transportationtomorrowsurvey/
[5] http://www.ncr-trans-rcn.ca/surveys/o-d-survey/o-d-survey-2011/
[6] https://www.saskatoon.ca/sites/default/files/documents/transportation-utilities/transportation/planning/Attachment3%20Technical%20Report%20HTS_FollowUp_report.pdf
[7] http://www.translink.ca/en/Plans-and-Projects/Transportation-Surveys.aspx
[8] http://transportation.speakupwinnipeg.com/WATS-Final-Report-July2007.pdf
[9] file:///C:/Users/khandker-admin/Downloads/tp_2011regionaltravelsurvey_030712.pdf
[10] http://www.icpsr.umich.edu/icpsrweb/ICPSR/studies/34910

| New York and New Jersey Metropolitan Area | Regional Household Travel Survey[11] | 2010-2011 | 0.24% |
|---|---|---|---|
| Southeast Florida | Household Travel Survey[12] | 2007-2008 | 0.11% |
| State of California | California Household Travel Survey[13] | 2010-2012 | 0.34% |
| Utah State | Household Travel Survey[14] | 2012 | 1.0% |
| **Australia** | | | |
| Adelaide | Travel Survey[17] | 1999 | 1.4% |
| Brisbane | Travel Survey[15] | 2009 | 1.3% |
| Canberra | Travel Survey[17] | 1997 | 2.6% |
| Greater Melbourne Area | Victoria Integrated Survey of Travel and Activity[16] | 2012 | 0.35% (per year) |
| Hobart | Travel Survey[17] | 2008-2009 | 2.9% |
| Sydney Greater Metropolitan Area | Continuous Household Travel Survey[17] | 2015 | 0.3% (per year) |
| **Europe** | | | |
| France | National Transport and Travel Survey[18] | 2007-2008 | Less than 0.1% |
| Germany | Mobilitat in Deutschland (MiD)[19] | 2008 | Less than 0.1% |
| The Netherlands | Onderzoek Verplaatsingen in Nederland (OViN)[20] | 2011 | 0.26% |
| Spain | Movilia[21] | 2007 | 0.31% |
| Switzerland | Microcensus on Travel Behavior[22] | 2010 | 0.67% |
| **South America** | | | |
| City of Rosario, Argentina | Household Travel Survey[23] | 2002 | 3% |
| Greater Santiago Area | Origin-Destination Survey[24] | 2012-2013 | 1% |

The first observation worth noting is that different regions/countries have developed their own patterns of household travel survey sample sizes. It is evident that many regions which had been collecting regular household travel survey data are gradually moving towards (or at least experimenting with) continuous surveys. Large scale continuous travel surveys are normally small scale repeated cross-sectional surveys collected in an ongoing fashion, rather than once every 5 or 10 years. In either case, household travel survey sample size determination is an important concern. Even for continuous surveys, it is recommended to pool the ongoing surveys in large intervals (3 or 5 years) to form a large pseudo cross-sectional survey (Ampt and Ortuzar 2004).

---

[11] http://www.nymtc.org/project/surveys/survey2010_2011RTHS.html
[12] http://www.fsutmsonline.net/images/uploads/mtf-files/Southeast_Florida_Household_Travel_Survey_0205_2014.pdf
[13] http://www.dot.ca.gov/hq/tpp/offices/omsp/statewide_travel_analysis/files/CHTS_Final_Report_June_2013.pdf
[14] http://www.wfrc.org/new_wfrc/publications/Utah_FinalReport_130228.pdf
[15] Stopher et al (2011)
[16] http://economicdevelopment.vic.gov.au/transport/research-and-data/vista
[17] http://www.bts.nsw.gov.au/Statistics/Household-Travel-Survey/default.aspx#top
[18] http://www.insee.fr/en/methodes/default.asp?page=sources/ope-enq-transports-deplac-2007.htm
[19] http://mobilitaet-in-deutschland.de/02_MiD2008/index.htm
[20] http://www.cbs.nl/nlnl/menu/informatie/deelnemersenquetes/personen-huishoudens/ ovin/doel/default.htm
[21] http://www.fomento.gob.es/mfom /lang_castellano/estadisticas_y_p ublicaciones/informacion_estadis tica/movilidad
[22] Ohnmacht et al 2012
[23] Ortuzar (2004)
[24] http://datos.gob.cl/datasets/ver/31616

Among all regions, Canadian cities are pioneers in large household travel surveys. Among all Canadian cities, Toronto and Montreal have regular (5-year interval) cross-sectional household travel surveys with a sampling rate of over 4.5%. Montreal has piloted a continuous household travel survey from 2009 to 2012 with an annual sample size of 15,000 households. The continuous survey was introduced between two large cross-sectional surveys conducted in 2008 and 2013. Other Canadian cities also regularly conduct household travel surveys with sampling frames of 2% to 5% of households. The City of Calgary is currently piloting a continuous household travel survey of 1,500 households per year over a 2-year period. Almost all Canadian household travel surveys are predominantly telephone-based with some introducing a web-version of the telephone survey and small scale GPS applications. Vancouver had the smallest sampling rate of all Canadian cities (2.2%). The metro region has stated in the past that the objective of the survey is mainly for model calibration purposes. The 2008 Metro Vancouver report mentioned that for obtaining detailed travel statistics such as trip rates and mode shares a larger sampling rate will be required. Nonetheless, the magnitude of such a survey may be too large adding costs and complexity to the data collection process.

On the other hand, cities and regions in the US have moved to small scale household travel surveys since the 1990s, potentially influenced by Smith (1979) and Stopher (1982). Almost all household travel surveys in the US have a sampling rate of less than 1%. However, US surveys are more dynamic in adopting advanced technology, e.g. GPS. The 2010-2011 New York and New Jersey regional household travel survey used a 10% sub-sample of households to collect a wearable GPS-based travel diary data. Even though the sample size remains small, the subsample proved to bring socio-economic groups that otherwise would not have participated in the survey. Further, the GPS subsample allowed the New York Metropolitan Transport Council along with the North Jersey Transportation Planning Authority to calculate statistically reliable trip rates that would have otherwise been more difficult to determine using a relatively small sample size. Still, the survey report recognizes the fact that this sample size might be too thin for various travel segments. The 2010-2012 California household travel survey employed a 12% sub-sample for a wearable GPS based travel survey. The biggest travel survey in the US is the National Household Travel Survey (NHTS) with a sampling rate of approximately 1%. However, in many cases, such data alone are not considered sufficient for demand modelling and evidence-based transportation planning exercises. The California household travel survey for example conveyed difficulty in determining detailed observed travel patterns at the county and/or sub-county levels due to the small sample size. Other difficulties reported include the underrepresentation of certain socio-demographic groups.

Australian cities have been implementing both cross-sectional and continuous travel survey approaches (Ortuzar 2011). Due to the lack of proper statistics, it is difficult to approximate the sample sizes of Australian surveys. However, it is clear that Australian surveys favor small sample sizes (Stopher et al 2011). Nevertheless, Stopher et al (2011) have highlighted the lack of consistency among these surveys thus limiting the potential of fusing the numerous datasets into one large survey, which the authors listed as an objective of various Australian planning agencies. One of the most intriguing Australian household travel surveys is the Sydney Household Travel Survey. Prior to 1997, the Greater Sydney Area used to conduct large scale cross-sectional surveys every 10 years. Since then, the area has been running a continuous survey. The data is pooled every 3 years, where the total sample size equals that of the pre-1997 cross-sectional survey. Other areas, such as the Central Melbourne area, use a cross-sectional household travel survey. The region uses both a land line based interview (55% of total sample) and a road-side intercept approach (45% of total sample) for data collection.

The European continent has the most consistent national household travel surveys. Bonnel and Armoogum (2005) stated that national household travel surveys in Europe vary widely in terms of sample sizes. Further, the authors report that the sample size determination is not correlated with the size or the characteristics of the countries respected populations. One of the critical sampling issues of national surveys is that the sampling process follows a variant of the cluster sampling approach. Cluster sampling may leave out several sub-regions from the data collection process. Thus, spatial distribution of travel behaviour at the smaller metropolitan level may become difficult. It is imperative to mention that European countries also conduct region-wide surveys. However, access to such survey details may have not been available online, or was hindered due to the information being displayed in a language other than English.

Chile, specifically the city of Santiago, has been a global leader in travel surveys. The latest Santiago household travel survey is of around 1% of households in the region. Chile also has been experimenting with various approaches e.g. continuous surveys, use of GPS technology and panel surveys (Ortuzar 2004).

Overall, it is clear that there is no consensus on the selection of sample sizes for household travel surveys. There are, however, recommendations on moving to continuous surveys instead of one-off surveys, but the issue of sample size is never tackled. Lack of proper data due to the small sample sizes of household travel surveys in the US has presented an issue for many researchers due to their inability to investigate detailed disaggregate (at a zonal or sub-regional level) travel behaviour. Some regions in the US have put forward the claim that small sample sizes prevent the observation of detailed travel patterns at the county or sub-county levels, and under represent certain segments of the population (SEFTC 2014). It is important to recognize that the sampling rates of countries can not be compared with those of regions, and it is also inaccurate to compare the sampling rate of one region with that of another. Nevertheless, the intent of table 1 is to present the reader with a sense of the common survey design practices around the world.

This paper is mainly concerned with the Canadian practice, specifically that of the GTHA. The next section considers a large-sample travel survey for empirical investigation on representation of such surveys.

## 4. Empirical Investigation on the Representativeness of a Large Sample Household Travel Survey: The Case of the Transportation Tomorrow Survey (TTS)

The TTS in the GTHA is one of the largest (5%) and most regularly conducted (every 5 years since 1986) household travel surveys in North America. The TTS study area is composed of 30 municipalities in addition to the City of Toronto's 16 planning districts; the largest municipality in the GTHA. The TTS has also been extended to include several smaller municipalities outside the borders of the GTHA. The 2011-2012 TTS survey data was used to investigate the TTS representativeness of the various socio-economic characteristics of its population. Figure 1 presents the aggregate region-to-region peak-period trip matrix of the study area of the TTS (DMG 2015).

Within the GTHA, the City of Toronto is the largest urban area with an established Central Business District (CBD). Its neighbouring regions of Halton, York, Peel and Durham feature independent municipalities. These regions function more or less as suburbs for Toronto. However, the Hamilton region is farther away from the City of Toronto and is also an established urban area. Almost all Origin-Destination pairs of the City of Toronto, Peel Region, and Halton Region have more than 1,100 peak period trips between them. Hence, based on the findings of Smith (1979), a 4% sample for these areas should be sufficient to adequately model trip behavior. However, in the case of the City of Hamilton and the Region of Durham and York, the majority of O-D pairs have less than 1,100 trips in the peak period. If we consider peak period transit trips, then the numbers are even worse.

| From/To | City of Toronto | Region of Durham | Region of York | Region of Peel | Region of Halton | City of Hamilton | Region of Niagra | Region of Waterloo | City of Guelph | County of Wellington | Town of Orangeville | City of Barrie | County of Simcoe | City of Kawartha Lakes | City of Peterborough | County of Peterborough | City of Orillia | County of Dufferin | City of Brantford | County of Brant | Region Totals |
|---|---|---|---|---|---|---|---|---|---|---|---|---|---|---|---|---|---|---|---|---|---|
| City of Toronto | 510,000 | 7,300 | 62,600 | 47,900 | 5,000 | 900 | 400 | 600 | 200 | . | . | 300 | 400 | . | 100 | . | . | . | 100 | . | 635,800 |
| Region of Durham | 51,800 | 72,800 | 15,900 | 3,200 | 300 | 100 | . | 200 | . | . | . | . | 100 | 400 | 800 | 100 | . | . | . | . | 145,700 |
| Region of York | 123,600 | 4,300 | 127,800 | 20,500 | 1,600 | 300 | 100 | 400 | 100 | . | . | 700 | 1,800 | 100 | 100 | . | 100 | . | . | . | 281,500 |
| Region of Peel | 92,600 | 700 | 18,200 | 188,600 | 15,900 | 1,400 | 500 | 1,200 | 600 | 200 | 600 | 100 | 700 | . | . | . | 100 | 200 | 100 | . | 321,700 |
| Region of Halton | 28,000 | 200 | 2,800 | 37,700 | 52,300 | 7,900 | 800 | 2,000 | 700 | 400 | . | 100 | 100 | . | . | . | . | . | 200 | 100 | 133,300 |
| City of Hamilton | 4,600 | 100 | 600 | 6,100 | 21,000 | 72,700 | 2,600 | 2,000 | 500 | . | . | . | . | . | . | . | . | . | 1,600 | 400 | 112,200 |
| Region of Niagra | 1,300 | 100 | 100 | 1,000 | 3,800 | 6,300 | 75,900 | 100 | 100 | . | . | . | . | . | . | . | . | . | 100 | . | 88,800 |
| Region of Waterloo | 1,300 | 100 | 200 | 3,100 | 1,800 | 1,200 | 100 | 104,400 | 7,600 | 1,300 | . | . | 100 | . | . | . | . | . | 800 | 400 | 122,400 |
| City of Guelph | 800 | . | 200 | 1,400 | 1,500 | 400 | 100 | 3,400 | 21,000 | 1,700 | 100 | . | . | . | . | . | . | . | . | . | 30,600 |
| County of Wellington | 400 | . | 200 | 1,900 | 1,100 | 100 | . | 2,100 | 4,200 | 3,600 | 300 | . | . | . | . | . | . | . | . | . | 13,900 |
| Town of Orangeville | 400 | . | 400 | 3,400 | 200 | . | . | . | 100 | 100 | 2,100 | . | 100 | . | . | . | . | 600 | . | . | 7,400 |
| City of Barrie | 1,600 | 100 | 2,800 | 1,000 | 100 | . | . | . | 100 | . | . | 17,500 | 5,400 | . | . | . | 600 | . | . | . | 29,200 |
| County of Simcoe | 4,000 | 400 | 8,300 | 3,000 | 200 | 100 | . | 100 | . | 100 | 300 | 10,500 | 29,300 | . | . | . | 3,600 | 300 | . | . | 60,200 |
| City of Kawartha Lakes | 300 | 2,600 | 500 | 100 | . | . | . | . | . | . | . | 100 | 200 | 8,600 | 1,500 | 300 | 100 | . | . | . | 14,300 |
| City of Peterborough | 100 | 700 | . | . | . | . | . | . | . | . | . | . | . | 500 | 12,400 | 2,100 | . | . | . | . | 15,800 |
| County of Peterborough | 100 | 500 | . | . | . | . | . | . | . | . | . | . | . | 400 | 5,700 | 2,000 | . | . | . | . | 8,700 |
| City of Orillia | . | . | 100 | . | . | . | . | . | . | . | . | 500 | 1,200 | . | . | . | 3,400 | . | . | . | 5,200 |
| County of Dufferin | 300 | . | 200 | 2,000 | 200 | . | . | 100 | 100 | 200 | 1,400 | . | 400 | . | . | . | . | 1,700 | . | . | 6,600 |
| City of Brantford | 200 | . | 100 | 300 | 600 | 2,100 | 100 | 1,200 | 200 | . | . | . | . | . | . | . | . | . | 1,200 | 2,400 | 8,400 |
| County of Brant | 100 | . | . | 100 | 300 | 900 | . | 1,700 | 100 | . | . | . | . | . | . | . | . | . | 2,800 | 2,400 | 8,400 |
| Region Totals | 821,500 | 89,900 | 241,000 | 321,300 | 105,900 | 94,400 | 80,600 | 119,500 | 35,500 | 7,700 | 4,800 | 29,800 | 39,800 | 10,000 | 20,600 | 4,500 | 7,900 | 2,800 | 6,900 | 5,700 | 2,050,100 |

**Figure1: Peak Period Trip Matrix of 2011-2012 TTS**

Moreover, in order to further investigate how well the 5% TTS sample represents the whole population, the Root Mean Square Error (RMSE %) was used to estimate the error/bias between the 2011 TTS and the 2011 census. Kish (1965) as well as NCHRP (2008) recommended this method to estimate error/bias in surveys. It is important to note that bias could be the result of sampling, in addition to measurement error, coverage error and non-response.

$$Percent\ RMSE = \sqrt{\frac{1}{n_i}\sum_i^{n_i}\frac{1}{n_{ji}}\sum_j^{n_{ji}}\left(\frac{r_{ij}-s_{ij}}{r_{ij}}\right)^2} \times 100 \qquad (1)$$

Where:
$n_i$ is the number of variables
$n_{ji}$ is the number of category j in variable i
$r_{ij}$ is the reference value (in census) of variable i in category j
$s_{ij}$ is the sample value of variable i in category j

As it is clear in this equation, the higher the errors for a particular variable, the higher is its representation bias of the whole population. We selected the socio-economic and household specific variables that are common between 2011 TTS and the 2011 census, considering census data as a reference. The following variables were used to estimate the RMSE of 2011 TTS data:
- Number of males
- Number of females
- Number of employed people
- User of modes:
  - Private car driver; Private car passenger; Transit users; Pedestrians; Bicycle users and Other mode users
- Age groups:
  - Under 14 years; 14+, up to 24 years; 24+, up to 44 years; 44+, up to 64 years and 64+ years
- Household sizes:
  - 1 person; 2 persons; 3 persons; 4 to 5 persons and 6 or more persons

Figure 2 presents the results of 6 cities in the GTHA. The cities are Toronto, Hamilton, Mississauga (Peel Region), Brampton (Peel Region), Oshawa (Durham Region), and Markham (York Region). The majority of the RMSE is below 20% for both 2011-2012 and 1991 TTS. In other words, TTS data represents its target population with an 80% accuracy margin. Part of the 20% error margin is germane to its sampling frame (land line phone directory), which cannot be eliminated by simply increasing sample size. Results show that non-motorized modes and transit modal shares have a higher error percentage than private automobile use. Error dispersion is higher in 1991 for cities other than Toronto. This may be due the adoption of a differential sampling strategy in 1991. Since the 1991 census didn't capture modal share, it was not possible to assess the accuracy of the 1991 TTS data. Nevertheless, it seems that a 5% sample can produce data representing the target population with an 80% plus margin of error.

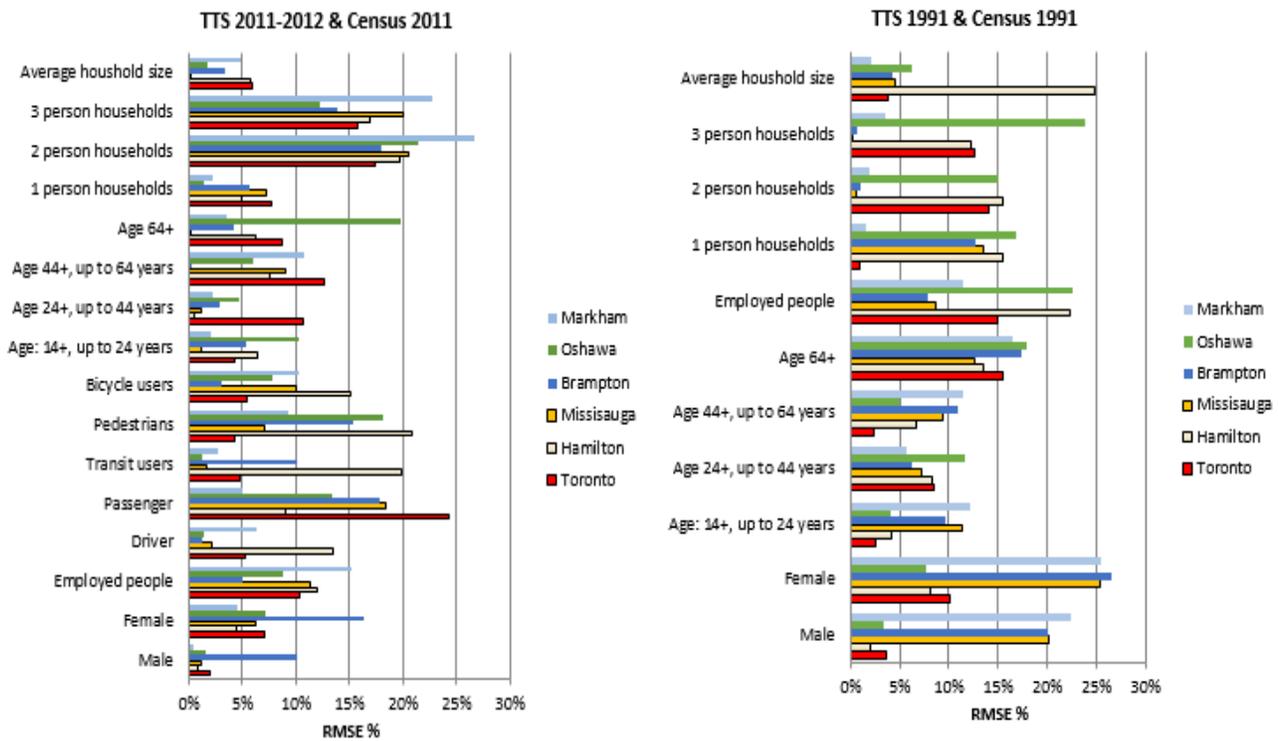

**Figure 2: RMSE of Selected Areas in the GTHA**

It is imperative to note, however, that the RMSE estimation of the cities and variables is dependent on variable availability, and the commonality of spatial boundaries in both TTS data and corresponding census data (Stats Can 2011). It is also important to note that the 2011 TTS featured a consistent sampling rate of approximately 5% across all regions. On the other hand, the 1991 TTS adopted a differential sampling rate distinguishing between "high growth" and "low growth" areas where the former was sampled at a 4.5% rate and the latter at 0.5%. The mean sampling rate of the 1991 TTS was 1.4% (DMG 2012).

## 5. Empirical Investigation on the Sample Size Requirement for Trip Distribution

It has been generally accepted that zone-to-zone trip matrices are difficult to construct with an appropriate margin of error using trips sampled via a household travel survey (TMIP, 1996). Cools et. al (2010) conducted an assessment of the quality of O-D trip matrices derived from activity surveys using a Monte Carlo experiment set up to estimate the precision of these matrices at various sampling rates. The authors calculated the Mean Absolute Percentage Error (MAPE) of OD matrices for different sampling rates generated using data from the 2001 Belgian national census. They concluded that only when half of the population is sampled can an acceptable O-D matrix be obtained at the provincial level, a sampling rate too large to be undertaken by any government authority. Nevertheless, the study also noted that an O-D matrix for peak period auto-mode commuter-only travel reproduced from a sampling rate of only 1% has a MAPE of 19%. Therefore, it is possible to construct statistically adequate O-D matrices if the level of disaggregation is not too thin.

As previously discussed, one of the most prominent pieces of work that related O-D trip matrices to sample size estimation is a graph proposed by Smith (1979). The graph represents the number of trips expected for a given interchange – from one spatial unit to another such as zone-to-zone or region-to-region. The rate however is based on randomly selected trips rather than randomly selected households. Smith argues that O-D trip matrices are simply not feasible because a high sampling rate is required to produce acceptable trip estimates. While not necessarily incorrect, this does not always have to be the case.

Smith pointed out that the sampling rate is correlated with the total number of trips between an origin and a destination. Nevertheless, he admitted that the relationship varied depending on the heteroscedasticity of the population, determined by the coefficient of variation of total trips. The coefficient of variation (CV) is a standardized statistical measure of dispersion of a frequency or probability distribution calculated by dividing the standard deviation by the mean (Searls 1964). Smith assumed a constant coefficient of variation of 1 when constructing his graph[25]. However, travel behaviour is not constrained by a specific distribution, rather it is best represented by a spectrum indicating potential homogeneous and heterogeneous travel patterns. Therefore, in an attempt to better understand the sample size requirements to construct O-D matrices, a portion of the graph was recreated using a range of CVs from 0.5 to 1.5 with a 0.25 increment.

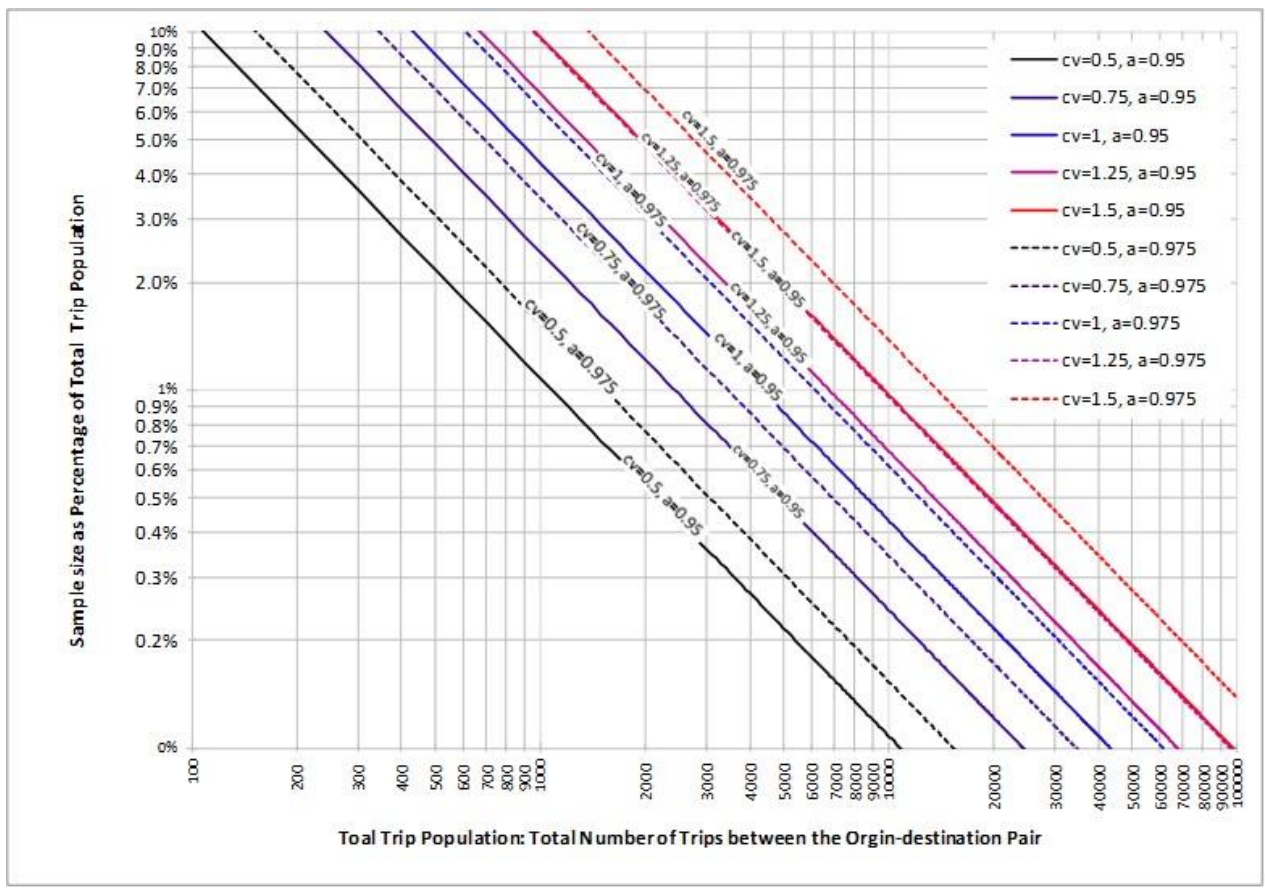

**Figure 3: Sample Rates for Trip Distribution Based on Trip Counts and CV**

---

[25] Smith (1979) does not explicitly state that a CV of 1 was used to construct the graph. Different points were selected from the graph to reverse calculate the CV, as the sampling rate, population (number of trips), confidence interval and margin of error were all provided.

The Y-axis of the graph represents the sample size (here, the number of trips between an O-D pair) to be surveyed divided by the trip totals (i.e. the sampling rate). The sample size is bounded by a confidence interval. It is calculated using formula 2 (page 19). Moreover, the number of trips to be captured will vary depending on the coefficient of variation, thus the different colors. A confidence interval of both 90% and 95% for trip totals between O-D pairs, along with a margin of error of 25%, were assumed. The X-axis is simply a series of hypothetical trip totals between O-D pairs. A logarithmic scale is assumed for both the y and x-axis.

The center blue line in figure 3 is the equivalent of the left-most solid line in Smith's graph. As can be observed in figure 3, as the CV decreases, the sample size requirements decrease accordingly and vice versa. Similarly, as the confidence interval increases from 90% to 95%, the required sampling rate also increases. Referring back to figure 1, the coefficient of variation calculated for the trip cells between Hamilton and the City of Toronto (i.e. the GTHA) is approximately 0.5. The plot shows that for a CV of 0.5 and O-D pairs with lower than 1,000 trips, a sampling rate of approximately 1% is required. Further, although some trip cell values are less than 1,000, many are in the order of 10,000. Thus, a 1% sampling rate may not be even necessary for such a spatially aggregated O-D matrix.

Cities that exhibit homogeneous travel behavior (e.g. auto-captive population) can have a similar CV, such as that reported by Pearson and others (TMIP, 1996)[26]. On the other hand, a multi-modal region with a number of residential and employment hubs like the City of Toronto has a CV of approximately 2, which will result in a larger sampling rate requirement[27]. Moreover, further disaggregation (by mode, peak and off-peak travel periods and trip purpose or spatial units) will require even larger sampling requirements, as the number of trips conducted from each origin to each destination will likely be smaller.

## 6. The Case of The TTS and Similar Large Scale Regional Travel Surveys

Household travel surveys used to be of large sample size. Even costly face-to-face household interviews in the early 1970s were at least 1% to 3% of population size (Stopher and Greaves 2007a). The increase of survey costs and the inherent complexity of political processes pertaining to transportation planning have generated different global approaches to household travel survey sizes. The practice of "small" (less than 1%) sampling rates was mostly motivated by Smith (1979). However, Canadian regions have been uniquely and consistently practicing the design and implementation of large scale household travel surveys. It seems that the practice of small scale surveys in the US is not always providing consistent and reasonable datasets for evidence-based planning, as provided in some of the survey reports discussed in section 3. With increasing under-coverage concerns related to the use of a land line based telephone directory, the telephone-only sampling frame and the increasing cost per completed responses, it has become imperative for the GTHA to identify a more adequate sampling rate as opposed to a common percentage of the population.

---

[26] Pearson reported in 1974 coefficient of variations of 0.53 for home-based work trip and 0.58 for home-based non-work trips.
[27] CV calculated for trips generated broken down by time of day, mode and trip purpose

Leveraging the lessons learned from the review conducted in this paper on sampling rates, and an empirical investigation conducted on previous TTS datasets, we propose a compound sampling approach that allows the regions/cities in the GTHA the flexibility to acquire sufficient data needed to maintain a good representation of the target population. The GTHA has been able to bear the cost of a 5% TTS survey over the last three decades. There is an intention to accommodate a similar percentage in the future, albeit increasing costs. However, it is necessary that an expensive survey as such should provide a representative sample while meeting the data needs of individual regions/cities within the GTHA.

We propose to follow a core-augment approach for survey sample size determination. The core refers to a common base sampling rate of 4% to develop a level ground for trips among the sub-regions of the whole study area. However, such a sampling rate will not be sufficient for the sub-regions with small population sizes as certain travel modes or segments of the community will be severely underrepresented. Augment samples should be used to complement the data needed for travel demand modelling; specifically, that of trip destination choice and travel mode choices for disaggregate activity-based models.

Having a common core sample is also important to build a general understanding of the market shares of different competing modes in the region, especially in the absence of a mandatory census long form (that collects household travel information from a representative sample of the whole population) in Canada. The recommendation of reducing the core sample from 5% to 4% is based on the understanding that an additional core of 1% does not provide a better representation than custom made region-specific augment samples.

Further, we recommend that the core 4% sample should be based on a stratified random sampling approach to maintain a good representation of population cohorts (age, gender, income, household car ownership, employment, occupation, etc.). Augment samples can be choice-based samples if specific mode users are under-represented, or attribute-based augmentations if general socio-economic cohort representation is of concern. The augment sample is intended to play one of two roles: to provide statistically reliable data on variables that were not considered in the core household survey; to compensate the underrepresentation of certain socio-demographic or trip-characteristic variables that are already present in the core survey. Examples of the latter objective for the TTS include the underrepresentation of young adults, and/or the thin public transit counts present in a number of smaller sized planning districts in the GTHA.

Different regions/cities within GTHA can set their priorities of transportation planning processes to define their augment samples (e.g. additional sample of transit riders, non-motorize mode users, low-income people, no-car households, students, empty nesters, etc.). This will provide regions/cities within the GTHA the flexibility in meeting their individual data needs, in addition to having a common database across the GTHA. The approach of Smith (1979) and Ampt and Ortuzar (2007) are recommended for the augment sample. An empirical exercise is presented in the following section.

## 7. Empirical Investigation on the Calculation of Sample Size for an Augment Sample

Smith (1979) proposed a heuristic to calculate sample size for household travel surveys. Smith's approach was regenerated by Stopher (1982) for different area zones in Michigan for a population of 1.6 million households, only to conclude that the approach is adequate to estimate sample size for trip generation purposes. Ortuzar (2004) took sample size estimation one step further by developing an optimization strategy to calculate sample sizes required for trip generation estimation of household surveys. His strategy, a multi-stage stratified random sampling technique, calculates an even smaller sample size through ordering socio-demographic variables by zone and subsequently selecting a random sample from these zones. Smith's approach was chosen over Ortuzar's advancement however due to its simplicity, permitting reproducibility among the different planning agencies. Further, Ortuzar's optimization of size and number of zones sampled may not be of great benefit to organizers with the objective of designing an attribute (versus spatial) based augment sample.

Consequently, leveraging the simplicity and adequate accuracy of the Smith's method, the sample size for two major cities (City of Toronto and Region of Hamilton) was calculated to provide an example for the proposed augmented sample recommended in this paper. The variables were chosen arbitrarily for the sake of demonstrating the calculation method and the expected results of an augment sample.

The 2011 TTS data was used for the estimation process. Households in each city were stratified in accordance with three major criteria: household size, income status and number of vehicles. This is an addition to the original Smith approach, which only focused on two variables. The planning organization of interest can further expand the number and vary the selection of variables as per their pre-determined objectives. Further, a 95% confidence interval was adopted along with a 5% acceptable error margin. Following the stratification, the sample size for each class was calculated. The sample sizes were then summed up to provide the total number of households required to estimate statistically reliable trip generation rates.

The following steps outline Smith's method for estimating the augment sample:

1- Stratify survey or trip data in accordance to individual or household socio-demographic or trip attributes.
2- Select variables of interest and designate every category representing a possible combination of the selected variables a number.
3- Calculate average trip rate per category.
4- Calculate the standard deviation of the average trip rate for each category.
5- Determine the CV per category by dividing the standard of deviation by the overall average trip rate.
6- Estimate the frequency of each category within the sample.
7- Normalize the CV by multiplying it with the category frequency.
8- Sum up step 7 to determine $C^*$
9- Choose a desired margin of error (e.g. 5%) and a corresponding confidence interval (e.g. 95%).
10- Calculate the initial sample size using the following equation:

$$F = C^{*2} \frac{Z^2}{E^2} \qquad (2)$$

Where,
- F is the initial sample size
- Z is a normal variate
- E is the margin of error

11- Divide the categorical factors calculated in step 7 by their sum $C^*$ to determine the weight of every category.
12- Multiply the weight of every category by the initial sample size calculated using formula number (2) to determine the optimal allocation of sample size by category.
13- Multiply each category frequency by the initial sample size as well to determine the expected frequency of the sample to be surveyed.
14- Identify the critical category in which the sample has the highest shortfall in estimating sample size. This may be determined by comparing the CVs and identifying the one with the largest value.
15- Divide the optimally allocated sample (step 12) of the critical category by its expected frequency sample distribution (step 13). This identifies the shortfall percentage.
16- Multiply the expected frequency sample distribution by the percentage increase identified in step 15 to calculate the required representative full random sample.

The resulting augment sample size for the city of Toronto was 5,344 households, which is around 0.5% of Toronto's household population. This number is more than fourfold the recommended sample size by Smith. This is mainly due to the stricter confidence interval and associated margin of error. Nonetheless, Toronto is a vibrant mobility hub with a population of 2.6 million of various socio-demographics affecting trip behaviour, and is a central city to its neighbouring region suburbs. Thus, a larger sample size may be required to capture the variation between the pre-specified classes. On the other hand, the total sample size required for Hamilton is 1,255 households, which is around 0.6% of Hamilton's population. This is equivalent to the upper end of sample sizes suggested by Smith.

One important issue is that this approach requires stratification and reference data. In case of household travel surveys, census data are normally used for the reference. However, in absence of census data only geographic stratification would be possible.

## 8. Empirical Investigation on the Calculation of Sample Size for the TTS Based on Trips by Time, Mode and Purpose

It has been previously mentioned in this paper that a trip distribution table in its most aggregate form can be broken down by time of day, mode and trip purpose. It comes as no surprise that obtaining more detailed trip distribution tables will come at the cost of increasing the overall sample size of the household travel survey.

To further investigate this matter, Smith's heuristic was leveraged to calculate the required sample size. The household non-expanded trips, extracted from the TTS database, were aggregated at two time points of a typical travel day, the first being the peak period between 6:00 am to 10:00 am, and 3:00 pm to 7:00 pm, while the latter being everything else. In the case of mode, trip counts were aggregated into four separate groups: automobiles, public transit, active modes (such as walking and cycling), and another category where all other potential modes were assigned. Finally, five trip destinations were created: work, home, school, recreational trips (including shopping) and an "other" category for any additional trip purposes not accounted for in the initial four categories.

A total of seven different classification tables using one, two or all three of the variables were produced. Within each classification table, the mean and standard deviation of the total number of household trips generated were calculated for each sub-category. Next, the cell frequencies were determined. Further, the modified CV for each subcategory and the required sample size for each classification table were then estimated from using Smith's method (Smith 1979), thus enabling the calculation of the required sampling rate. The results show an increasing sampling rate trend from 1% and 8% needed to accurately capture trip distribution for the TTS - dependent on the level of disaggregation.

**Table 2: Sample Rates for Trip Distribution at Various Disaggregation Levels of Time of Day, Mode, and Trip Purpose**

| Disaggregation Level | Sample Rate |
|---|---|
| By mode | 1% of households |
| by purpose | 4% of households |
| by time | 4% of households |
| by time & mode | 4% of households |
| by time & purpose | 5% of households |
| by purpose & mode | 8% of households |
| by time, purpose & mode | 8% of households |

## 9. Conclusion and Suggestion for Further Research

This paper investigates the issue of sample size determination for household travel surveys. An extensive review of existing literature revealed different global practices. Only Canadian regions have been able to maintain large sample household travel surveys while most other countries have chosen a smaller sampling rate, or adopted new approaches like continuous surveys. Toronto and Montreal are prime examples of cities implementing large cross-sectional surveys. While the move towards small sample household travel surveys is mainly driven by budget limitations, theoretical justification was not necessarily neglected. However, although small sample sizes are theoretically acceptable, the approach often fails to provide sufficient data for long-term trend analysis and disaggregate travel demand modelling. The Canadian examples have proven that, even with the increasing cost of implementing surveys, it is possible to maintain large sample household travel surveys.

Nevertheless, it has been also proven that a large sample size survey does not necessarily equate to a representative survey. Empirical investigation revealed that even the TTS with a sample size of more than 150,000 households (a 5% sample) can have an error of over 15% in representing basic population cohorts and attributes. Further, this study indicated that the sample size requirements for constructing a statistically adequate O-D matrix are proportional to the CV ratio of the population. The sample size also increases if the intent of the respective region or country is to capture O-D pair trip counts at a more disaggregate level (e.g. by mode, time of day, trip purpose, etc.)

Therefore, the paper proposes a core-augment approach of sample size determination for household travel surveys. In the case of the GTHA, the proposed approach suggests a core sampling rate of 4% common to all regions/cities of the study area. Further, additional regions/cities can implement custom-made augment surveys specific to their data needs. It is recommended that the core common sample follows a stratified random approach to have a good representation of the whole population of the study area. The size of the core sample should be statistically able of adequate (e.g. defined acceptable errors and confidence limits) population representation while factoring budget limitations.

The study however does not come without limitations. The effect of stratified sampling (or other forms of sampling for that matter) on sample size has not been accounted for. In addition, although It has been established that the CV is a major determinant of sample size, limited effort has been invested in understanding the variance exhibited in different forms of travel behaviour. Moreover, the relationship between the different types of bias and sample size was not expanded on. Finally, the study was contextual to the survey needs of the GTHA and similar regions in Canada.

This paper investigated solely the issue of sample size requirements of household travel surveys without necessarily considering the issues of survey cost, sampling frame and continuous versus one-off cross-sectional survey choice. Identifying an appropriate sampling frame is another critical factor that can inhibit the representation of large scale household travel surveys. It is necessary to investigate whether any innovative or hybrid sample frame and/or survey mode (e.g. smart phone, GPS, etc.) can further reduce the base cost of household travel surveys.
While budget limitations are an unavoidable reality, it is important to investigate the direct and indirect benefits of large scale household travel surveys, including potential future money savings from limiting the implementation of inefficient infrastructure investments. Such savings can offset and justify the high cost of large scale surveys.

## Acknowledgement

This study was funded by the Transportation Information and Systems Committee (TISC) representing the planning agencies in the GTHA and headed by the MTO. Authors acknowledge the comments and suggestions of Eric Miller, Siva Srikukenthiran, Patrick Loa, Mohammed Rashed and Shashank Pulikanti. However, the views and opinions presented in the paper belong only to the authors.

## References

Ampt, E.S., Ortuzar, J.D. 2004. On best practice in continuous large-scale mobility surveys. Transport Reviews 24(3): 337-363
Badoe, D.A., Steuart, G.N. 2002. Impact if interviewing by proxy in travel surveys conducted by telephone. Journal of Advanced Transportation 36(1): 43-62

Bolbol, A., Cheng, T., Tsapakis, I., Chow, A. 2012. Sample size calculation for studying transportation modes from GPS data. Procedia-Social and Behavioural Science 48: 3040-3050

Bonnel, P. and J. Armoogum (2005) National Transport Surveys – What can we learn from international comparisons? Paper presented at the European Transport Conference, Strasbourg, October.

Cools, M., Moons, E., Wets, G. 2010. Assessing the Quality of Origin-Destination Matrices Derived from Activity Travel Surveys. Results from a Monte Carlo Experiment. Transportation Research Record: Journal of Transportation Research Board 2183: 49-59.

Data Management Group: DMG. 2015. http://www.dmg.utoronto.ca/transportationtomorrowsurvey/

Fleiss, J. 2003. Statistical Methods for Rates and Proportions. Second Edition. John Wiley & Sons, New York.

Goulias, C., Pendyala, R., Bhat, C.R. 2013.Total design data needs for the new generation large scale activity microsimulation model. In *Transport Survey Methods: Best Practice for Decision Making*, J. Zamud, M. Lee-Gosselin, J.A. Carrasco and M.A. Munizaga (eds). Emareld Group Publishing, UK

Greaves, S. P. and Stopher, P.R. 2000. Creating a synthetic household travel and activity survey: Rationale and feasibility analysis. Transportation Research Record 1706: 82–91.

Harvey, A.S. 2003. Time-space diaries: merging traditions. In: Stopher, P., Jones, P. (Eds.), Transport Survey Quality and Innovation. Pergamon Press, pp. 151–180. http://dc.chass.utoronto.ca/cgi-bin/census/2011nhs/displayCensus.cgi?year=2011&geo=da

JRC. 2013. Analysis of National Travel Statistics in Europe. JRC Technical Report. https://ec.europa.eu/jrc/sites/default/files/tch-d2.1_final.pdf (accessed in July 2015)

Kish, L. 1965. Survey Sampling. John Wiley & Sons, New York.

Kollo, H.P.H., Purvis, C.L. 1984. Changes in regional travel characteristics in the San Francisco Bay Area: 1960-1981. Paper presented at 1984 Transportation Research Board Annual Meeting, Washington DC, January 1984

National Cooperative Highway Research Program: NCHRP. 2008. Standardized procedure for personal travel surveys. Transportation Research Board. Washington DC

Ohnmacht, R., Rebmann, K., Brugger, A. 2012. Swiss microcensus on mobility and transport. Paper presented at the 12[th] Swiss Transport Research Conference, Monte verita, May 02-04, 2012

Ortuzar, J.D. 2004. Travel survey methods in Latin America. Paper presented at the 7[th] International Conference on Survey Methods in Transport. Costa Rica 2004. http://www.isctsc.cl/archivos/2004/Keynote%20paper%20Ortuzar%20rev.pdf (Accessed in July 2015)

Ortuzar, J.D., Armoogum, J., Madre, J.L., Potier, F. 2011. Continuous mobility surveys: The state of practice. Transport Reviews 31(3): 293-312

Pointer, G., Stopher, P.R., Bullock, P. 2004. Monte Carlo simulation of household travel survey data for Sydney, Australia: Bayesian updating using different local sample sizes. Transportation Research Record 1870:102-108

Richardson A. J., Ampt, E. S. and Meyburg, A. 1995. Survey Methods for Transport Planning (Melbourne: Eucalyptus).

Searls, D.T. 1964. The Utilization of a known Coefficient of Variation in The Estimation Procedure. Journal of American Statistical Association 308(59): 1225-1226

Smith, M.E. 1979. Design of small sample home interview travel surveys. Transportation Research Record 701: 29-35

Southeast Florida Transportation Council: SEFTC. Southeast Florida Household Travel Survey. 2014. Modeling and Regional Transportation Technical Advisory Committee.

Statistics Canada: StatsCan. Census of Canada. 2011. From Canadian Census Analyser.

Stopher, P. 1982. Small sample home-interview travel surveys: Application and suggested modification. Transportation Research Record 886: 41-47

Stopher, P. Zhang, Y., Armoogum, J., Madre, J-L. 2011. National household travel surveys: The case for Australia. Proceedings of 2011 Australian Transport Research Forum Conference, 28-30 September 2011, Adelaide, Australia.

Stopher, P., Greaves, S.P., 2007a. Household travel surveys: Where are we going? Transportation Research Part A 41: 367-381

Stopher, P., Greaves, S.P., 2007b. Guidelines for samples: measuring a change in behaviour from before and after survey. Transportation 34:1-16

Stopher, P., Kockelman, K., Greaves, S.P., Clifford, E. 2008. Reducing burden and sample sizes in multi-day household travel surveys. Paper presented at the 87th Annual Meeting of the Transportation Research Baord. Washington DC, January 2008

Stopher, P., Meyburg, A.H. 1979. Survey sampling and multivariate analysis for social scientists and engineers. Heath, Lexinton, MA, USA

TransLink. TransLink's 2008 Regional Trip Diary Survey: Final Report. 2010. South Coast British Columbia Transportation Authority.

Travel Model Improvement Program: TMIP. 1996. Travel Survey Manual. Prepared by Cambridge Systematic Inc. for US DOT, FTA, FHA, OoC and US EPA. www.fsutmsonline.net/images/uploads/mtf-files/Southeast_Florida_Household_Travel_Survey_0205_2014.pdf